\documentclass[prl,twocolumn,nofootinbib, preprintnumbers, superscriptaddress]{revtex4}

\usepackage{amsmath,amssymb,slashed,braket}
\usepackage{graphicx}
\usepackage{epstopdf}
\usepackage{float,appendix}
\usepackage[colorlinks=true,
            linkcolor=blue,
            urlcolor=blue,
            citecolor=green,          
            bookmarks=true,
            bookmarksnumbered=true,
            breaklinks=true,
            pdfpagemode=Fullscreen,
            pdfstartview=FitBH]{hyperref}

\usepackage{esint}

\usepackage[normalem]{ulem}
\usepackage{color}

\definecolor{Orange}{cmyk}{0,0.61,0.87,0}
\definecolor{JungleGreen}{cmyk}{0.99,0,0.52,0}
\definecolor{OliveGreen}{cmyk}{0.64,0,0.95,0.40}
\definecolor{Brown}{cmyk}{0,0.81,1,0.60}
\definecolor{RoyalBlue}{cmyk}{0.71,0.53,0,0.12}
\definecolor{Gray}{cmyk}{0,0,0,0.40}
\definecolor{LightPink}{cmyk}{0.0,0.25,0,0}
\definecolor{LLightPink}{cmyk}{0.0,0.10,0,0}
\definecolor{LightBlue}{cmyk}{0.25,0,0,0}
\definecolor{LightGray}{cmyk}{0,0,0,0.2}

\usepackage{xcolor}
\definecolor{gesfpurple}{rgb}{0.47,0.19,0.42}

\definecolor{gesflanse}{rgb}{0.00,0.50,0.50}

\definecolor{gesfblue}{rgb}{0.08,0.42,0.76}

\definecolor{gesfred}{rgb}{1,0,0}

\definecolor{gesfwhite}{rgb}{1,1,1}

\definecolor{gesfblack}{rgb}{0,0,0}

\newcommand{\geqn}[1]{Eq.\,\hypersetup{linkcolor=blue}(\ref{#1})\hypersetup{linkcolor=blue}}
\newcommand{\gfig}[1]{{\hypersetup{linkcolor=violet}Fig.\,\ref{#1}\hypersetup{linkcolor=blue}}}
\newcommand{\gtab}[1]{{\hypersetup{linkcolor=gesflanse}Tab.\,\ref{#1}\hypersetup{linkcolor=blue}}}

\graphicspath{{figs/}}

\begin{document}

\title{Detecting a Fifth-Force Gauge Boson via Superconducting Josephson Junctions}

\author{Yu Cheng}
\email{chengyu@sjtu.edu.cn}
\affiliation{Tsung-Dao Lee Institute \& School of Physics and Astronomy, Shanghai Jiao Tong University, China}
\affiliation{Key Laboratory for Particle Astrophysics and Cosmology (MOE) \& Shanghai Key Laboratory for Particle Physics and Cosmology, Shanghai Jiao Tong University, Shanghai 200240, China}

\author{Jie Sheng}
\email[Corresponding Author: ]
{shengjie04@sjtu.edu.cn}
\affiliation{Tsung-Dao Lee Institute \& School of Physics and Astronomy, Shanghai Jiao Tong University, China}
\affiliation{Key Laboratory for Particle Astrophysics and Cosmology (MOE) \& Shanghai Key Laboratory for Particle Physics and Cosmology, Shanghai Jiao Tong University, Shanghai 200240, China}

\author{Tsutomu T. Yanagida}
\email{tsutomu.tyanagida@sjtu.edu.cn}
\affiliation{
Kavli IPMU (WPI), UTIAS, University of Tokyo, Kashiwa, 277-8583, Japan}
\affiliation{Tsung-Dao Lee Institute \& School of Physics and Astronomy, Shanghai Jiao Tong University, China}
\affiliation{Key Laboratory for Particle Astrophysics and Cosmology (MOE) \& Shanghai Key Laboratory for Particle Physics and Cosmology, Shanghai Jiao Tong University, Shanghai 200240, China}

\begin{abstract}

A new fifth force between particles carrying $B-L$ charges is well-motivated by the intriguing $U(1)_{B-L}$ extension of the standard model. The gauge boson mediator, F\'eeton, also serves as a dark matter candidate. In this letter, we propose a novel experimental design to detect the quantum phase difference caused by this fifth force using a superconducting Josephson junction. We find that the experiment has the best sensitivity to the gauge coupling when the gauge boson is within the mass range of $0.01\,$eV to $10\,$eV, which is an interesting mass region for the F\'eeton dark matter. This opens up a new avenue for the measurement of new physics at small scale below millimeter.

\end{abstract}

\maketitle 

{\bf Introduction}--
More than 80\% of the matter in our Universe today 
is dark matter (DM) \cite{Arbey:2021gdg}. But the nature of DM is still
a mystery and suggests new physics beyond the Standard
Model (SM) \cite{Young:2016ala}. 
The $B-L$ extension of the SM is a quite appealing option.
It requires three right-handed neutrinos (RHNs) to cancel the gauge anomalies. 
The presence of the heavy RHNs is
a key point for the natural explanation of the observed small neutrino masses through the seesaw mechanism \cite{Minkowski:1977sc, Yanagida:1979as, Yanagida:1979gs, GellMann:1980vs}, as well as for 
the creation of the baryon asymmetry in the present Universe via the leptogenesis \cite{Fukugita:1986hr, Buchmuller:2005eh}. 
The new $U(1)_{B-L}$ symmetry also predicts a new particle, the $B-L$ gauge boson (F\'eeton).
If the gauge coupling constant is extremely small,  
F\'eeton becomes a candidate for dark matter\cite{Choi:2020kch,Okada:2020evk,Lin:2022xbu} which can also generate a new fifth force by its exchange between matters.
These low-energy predictions offer numerous phenomena that can be tested.

\begin{figure}[!t]
\centering
 \includegraphics[width=0.47
 \textwidth]{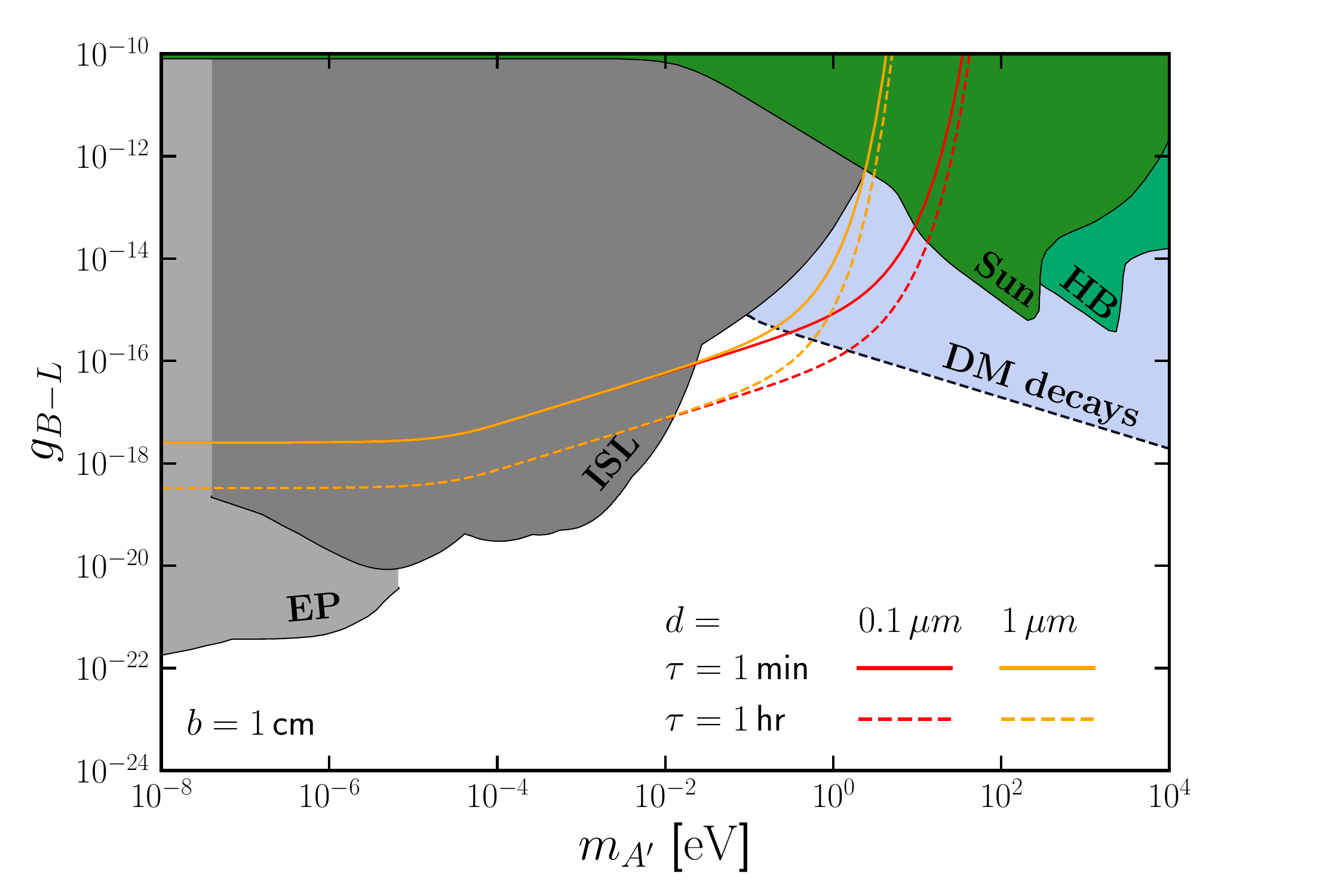}
\caption{ The shaded regions
are current constraints of $B-L$ 
gauge interaction from Torsion-balance experiments (EP, ISL), stellar cooling bounds from the Sun and Horizontal 
branch stars (HB). Assuming the gauge boson to be DM, 
the stability condition 
($\Gamma_{A^\prime \rightarrow \nu \bar \nu} 
< H_0$) exclude the light blue shaded region by taking the lightest neutrino mass to be zero and normal hierarchy.
The red ($d = 0.1\,\mu$m) and 
orange ($d = 1\,\mu$m) lines are the 
projected limits of our experimental proposal. The different line types, dashed and solid, are for different 
running times of one minute and one hour.}
\label{fig:limit}
\end{figure}

The test of low mass $B-L$ gauge boson is based on the fifth force detection.
For F\'eeton masses smaller than $10^{-2}\,$eV, the gauge force exhibits long-range behavior and receives strong constraints from gravitational wave detection \cite{LIGOScientific:2021ffg,Miller:2023kkd,Frerick:2023xnf} and torsion-balance experiments \cite{Wagner:2012ui,Adelberger:2009zz}.
However, as the mediator mass exceeds $10^{-2}\,$eV, the force range decreases, and its strength is greatly suppressed within the same distance. The constraint from torsion balance experiments becomes weaker, as depicted in Figure \ref{fig:limit}.
Due to the challenges associated with detecting a faint signal on a small scale, there are currently no terrestrial direct detection experiments specifically targeting the fifth force with mediator masses ranging from $0.01\,$eV to $10\,$eV. 
The purpose of this paper is to propose a novel experiment method to measure the F\'eeton fifth force in this particular mass region.

The experimental setup proposed consists of a Josephson junction circuit and a $B-L$ gauge potential source.
Because of the distance dependence of 
the $B-L$ gauge potential, the two superconductors inside a junction feel different fifth force strengths
from the source. 
Consequently, the two quantum superconducting 
states will evolve to have a phase difference
after a period of time \cite{Ummarino:2020loo,Perez:2023tld}. Such a phase difference
can be converted into a detectable charge current
due to the Josephson effects.
We first give a general derivation of the 
phase difference induced by the F\'eeton fifth force. 
Then, the experimental design and 
the predicted current are presented. After considering all the backgrounds, we finally 
establish the projected constraint on the $B-L$ gauge
coupling strength.

{\bf Phase Difference Induced by 
$B-L$ Gauge Interaction} -- 
In this paper, we consider the simple extension of SM with a $U(1)_{B-L}$ gauge symmetry. The detailed definition of the model is given in the Appendix.

The model leads to a new gauge force mediated by the $B-L$ gauge boson, F\'eeton.
In the non-relativistic regime, such a vector interaction can be written in 
terms of the Yukawa type of potential function, 
\begin{equation}
    V_{B-L}(r) = g_{B-L}^2 \frac{Q_{B-L} e^{-m_{A'} r}}{r}.
\end{equation}
%
The $Q_{B-L}$ is the product of the corresponding $B-L$ charges of two interacting bodies.

\begin{figure}[!t]
\centering
 \includegraphics[width=0.47
 \textwidth]{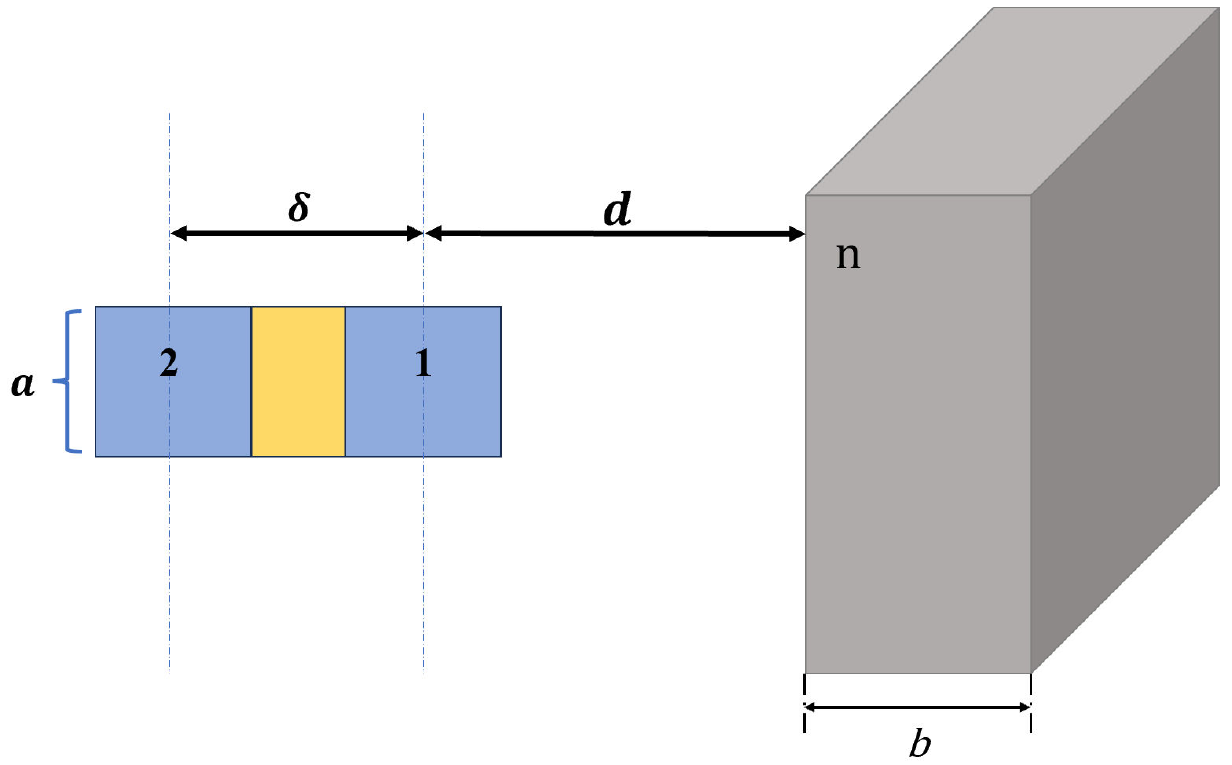}
\caption{ Two configurations of coherent states $\ket{1}$ and $\ket{2}$ 
with sizes $a$ are 
in different positions with 
a separation of $\delta$. 
A large plate with a thickness $b$ made of neutral atoms
is placed at a distance of $d$.
It carries the neutron
($B-L$ charge) 
density $\rho$ and serves as a 
potential source.
The above four length parameters have the relationship as $a \sim \delta \lesssim d < b$. 
}
\label{fig:PhaseDifference}
\end{figure}

According to the Schrodinger equation, 
$i \partial_t \psi = \hat H \psi$, the quantum phase $\phi (r, \tau)$ of any stationary 
state feeling the 
$B-L$ gauge force 
will evolve as 
$ i \phi (r, \tau) 
= i \int_0^\tau V_{B-L} (r) dt$. Since the potential is distance-dependent, two states in different positions will acquire a phase difference over a period of evolution.
This quantum phase effect
can be detectable and utilized to probe this tiny F\'eeton fifth force.
The settings are illustrated 
in \gfig{fig:PhaseDifference}.

 One first prepares two groups 
of particles in coherent states, $\ket{1}$ and $\ket{2}$\footnote{They can be some coherent particles splitted into a superposition of positions, atoms in Bose-Einstein condensation, Cooper pairs in superconductor, and so on. Our work focuses on the case of Cooper pairs.}.
Both states have a configuration with size $a$,  $B-L$ charge $Q_{B-L}$, 
and are in different
space positions with a small separation of $\delta$.
At the distance of $d$ \footnote{
In principle, particles at different positions within the same coherent state feel different external potential from the plate. 
However, they undergo the same phase in time evolution, i.e., the Cooper pairs have the same phase $\phi_1$ or $\phi_2$ as shown in Eq.(5) inside the superconductor 1 or 2, respectively. 
This is ensured by a small variation in the number density of Cooper pairs in superconductor, as the gradient of the phase implies the generation of a current.
Therefore, the phase of the coherent state should be an average of the phases of particles at different positions, effectively representing the phase at the midpoint of the superconductor. Thus, the distance $d $ means the average distance of superconductor 1 from the surface of the plate. The separation 
$\delta$ means the average distance between superconductor 1 and 2.}, 
there is a plate (a thick film) made of neutral atoms, which has a thickness of $b$ and its length size is large enough to be approximated as infinity. The above four 
lengths have the following hierarchy, $b > d \gtrsim \delta \sim a$.
While the $B-L$ charges of protons and electrons inside an atom cancel each other out, the neutrons contribute to a non-zero $B-L$ charge.
The neutron number density of the plate is $n$. 
With the integration of 
contributions from the 
whole volume, 
the phase of $\ket{1}$ induced by $B-L$ gauge potential with a distance $d$ is,
\begin{align}
\hspace{-2mm}
    \phi_f (d)
& =
    \tau g_{B-L}^2 Q_{B-L} n 
   \int_0^b \int_0^\infty 
   \frac{e^{-m_{A'} \sqrt{(d+z)^2 + r^2}} 2 \pi r}{\sqrt{(d+z)^2 + r^2}}
   d r dz
\nonumber\\
& =
    \tau g_{B-L}^2 Q_{B-L} n
    \frac{2 \pi e^{-m_{A'} d} \left(
    1-e^{-m_{A'} b}
    \right) }{m^2_{A'}}.
\end{align}
for a time period $\tau$.
Here, index $f$ labels for the fifth force.
For $\ket{2}$, the distance argument is $d+\delta$.  
Because of the position difference,  the $B-L$ gauge interaction sourced from the plate induces a phase 
difference $\Delta \phi_{f}$ between
states $\ket{1}$ and $\ket 2$ as,
\begin{equation}
    \Delta \phi_{f}
\equiv   \phi_f (d) - \phi_f (d + \delta),  
\label{phaseDef}
\end{equation}

In the limit of tiny separation $\delta < d$
and long range force $d \ll 1/m_{A'}$,
$e^{-m_{A'} (d+\delta)}
\simeq e^{-m_{A'} d} (1 - m_{A'} \delta)
\simeq (1 - m_{A'} \delta)$.
To maximize the signal, the thickness of the plate should be 
large enough that, $e^{-m_{A'}b} \sim 0$.
Thus, the \geqn{phaseDef} becomes,
\begin{equation}
    \Delta \phi_f 
\simeq 
    \frac{2 \pi  \tau g_{B-L}^2 Q_{B-L} n
 \delta}{m_{A'}} .
\label{phaseDiff}
\end{equation}
With the experimental materials determined, 
the phase difference can be enhanced by 
adjusting
the running time $\tau$ and 
the fifth force range $1/m_{A'}$. 
A longer force range makes the quantum states
feel more $B-L$ charges in plate,
and a larger phase 
is accumulated over an extended period of time.


{\bf Josephson Junction Detector} --
We have seen that the $B-L$ gauge interaction induces a phase difference
between two groups of coherent states in 
different space positions. 
The superconducting Josephson Junction (JJ) \cite{Josephson:1962zz,Josephson:1974uf,Ummarino:2020loo,Perez:2023tld} can transfer such a phase 
difference to charge current which can be detectable if it is sufficiently large.

A JJ is made of two separated superconductors 
and an insulator with width $\epsilon$
in between. Once the temperature of 
the superconductor falls below its critical temperature, 
the electrons inside the conductor
form Cooper pairs, becoming in a coherent state,
and the resistance disappears.
These Cooper pair states in the two superconductors are the coherent states
$\ket 1$ and $\ket 2$ as we want.
They can be described by order parameters \cite{Landau:1950lwq}, 
\begin{equation}
    \Psi_1 = \sqrt{n_1} e^{i \phi_1}
\,,
\quad 
    \Psi_2 = \sqrt{n_2} e^{i \phi_2},
\label{SCwave}
\end{equation}
analogous to a quantum mechanical wave function.
Here, $n_i$ ($i = 1, 2$) is the number 
density over the volume of Cooper pairs and $\phi_i$ is 
the corresponding phase. For simplicity, 
we take the two superconductors the same 
and $n_1 = n_2$. From the semi-classical
London theory, one can estimate the number 
density 
$n_e = m_e / (e^2 \lambda_L^2) $ through
the London penetration length $\lambda_L$ \cite{London:1935}. Usually, the penetration 
length varies from $50 \sim 500\,$nm for 
different types of materials. We take 
$\lambda_L = 50\,$nm so that the 
Cooper pair number density is $n_1 = n_2 = 1.2 \times 10^{22}\,$cm$^{-3}$.

In the insulator region, 
there is a barrier potential $V$ higher
than the kinetic energy of electrons $E$.
The wave functions can be parameterized as \cite{gross2016applied}, 
\begin{equation}
    \Psi (x)
=
C_1     \cosh{x/\xi} + 
C_2    \sinh{x/\xi},
\end{equation}
with $\xi = \sqrt{1/ 4 m_e (V - E)}$.
Taking \geqn{SCwave} as boundary conditions, one can solve, 
\begin{equation}
\hspace{-3 mm}
C_1 =
    \frac{\sqrt{n_1} e^{i \phi_1} 
    + \sqrt{n_2} e^{i\phi_2} }{2 
    \cosh (\epsilon/2\xi) }\,,
C_2 =
    \frac{\sqrt{n_1} e^{i \phi_1} 
    - \sqrt{n_2} e^{i\phi_2} }{2 
    \sinh (\epsilon/2\xi) }.
\end{equation}
Although classically forbidden, the 
quantum tunneling effect induces
a current.
The current density is generally 
$J = - (2 e/m_e) {\rm Re} (\Psi^* i \nabla \Psi)$\footnote{
The gauge invariant current should contain the gauge field $\mathbf{A}$
as $J = - (2 e/m_e) {\rm Re} [\Psi^* (i \nabla - e\mathbf{A}) \Psi]$.
In our setup, the possible background electromagnetic field is shield as discussed below to ensure $\mathbf{E} = \mathbf{B} = 0$. Thus, we can take the gauge $\mathbf{A} = 0$ and neglect this term in calculation.
}, and proportional to 
the phase difference as \cite{Terry:1901}, 
\begin{equation}
    J = \frac{e \sqrt{n_1 n_2}}{m_e \epsilon} \sin \Delta \phi\,, 
\quad 
\Delta \phi \equiv (\phi_1 - \phi_2).
\end{equation}

First we locate the plate far away from the JJ. 
In the beginning, the JJ
has an random initial phase difference, and some environmental background, such as Earth gravity, and the node connected to the JJ, can also generate Josephson current.
However, the current
can be dissipated by connecting a circuit 
with resistance and the junction goes 
into its ground state $\Delta \phi = 0$.
Then, we disconnect the circuit and put the plate at a close distance $d$ to the JJ.
The phase difference $\Delta 
\phi$ begins to evolve due to the $B-L$ gauge potential.
Finally at some moment $\tau$, we connect the circuit again to measure the electric Josephson current and compare the result with the theoretical prediction.
Considering the areal size of the cubic superconductor $a^2$, which should be larger than the penetration length and taken as $a \simeq 100\,$nm,
the current induced by the fifth force $I_f$ is, 
\begin{equation}
    I_f = I_c \sin \Delta \phi_f\,,
\quad 
\text{where}
\quad I_c = \frac{e n_1 a^2}{m_e \epsilon}.
\end{equation}

For illustration, we propose the 
plate material to be Graphene made of carbon atoms. 
From its mass density, $\rho_{C} = 2.27$\,g/cm$^{3}$,
one can estimate its neutron number density $n = 6.8 \times 10^{23}$\,cm$^{-3}$.
Inside a JJ, the insulator usually has a width $\epsilon = 1\,$nm and the Cooper-pair
quantum states have a $B-L$ charge $Q_{B-L} = -1 \times 2 = -2$ for two electrons.
Assuming the running time is one minute, $\tau = 60\,$s, 
plate thickness $b = 1\,$cm,
the induced phase difference \geqn{phaseDiff}, 
in the long-range-force limit,
is,
\begin{equation}
\hspace{-2mm}
    \Delta \phi_f 
=
    3 \times 10^{-3} \times 
    \left( 
    \frac{g_{B-L}}{10^{-16}}
    \right)^2
    \left( \frac{\tau}{1\,\text{min}}\right) 
    \left( \frac{10^{-2}\,\text{eV}}{m_{A'}}\right) ,
\label{dphi2}
\end{equation}

which is hopeful of being detected \cite{everitt2011gravity}. For a tiny phase difference, 
$\sin \Delta \phi_f \simeq \Delta \phi_f$. 
The corresponding Josephson current induced
from the fifth force is, 
\begin{align}
\hspace{-3mm}
    I_f 
& \simeq
     I_c \Delta \phi_f
\simeq 
    \frac{2 \pi e  \tau g_{B-L}^2 Q_{B-L} n n_1 a^2 b}{m_e}
\nonumber\\
& 
= 
6 \times 10^{-3}\,\text{A} \times 
\left( \frac{g_{B-L}}{10^{-16}}\right)^2 \left( \frac{\tau}{1\,\text{min}}\right) 
\left( \frac{10^{-2}\,\text{eV}}{m_{A'}}\right).
\label{currentFinal}
\end{align}
Again, with the experimental settings fixed, the signal is 
proportional to the running time, force range, and 
gauge coupling. As indicated by \geqn{dphi2}, 
The sine periodic variation of the current 
can even be observed for around one-hour running.
\begin{figure}[!t]
\centering
 \includegraphics[width=0.44
 \textwidth]{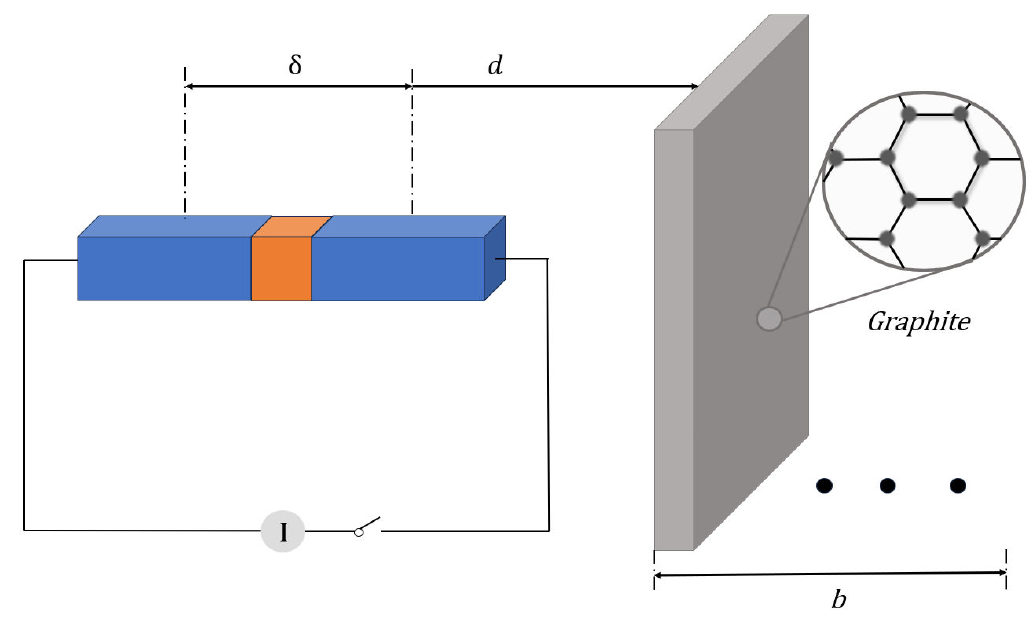}
\caption{ The experimental design for 
detecting the $B-L$ gauge interaction.
The Josephson junction
(with two superconductors as blue cube and one insulator as yellow cube)
is placed inside a conductor (gray shaded) with a 
symmetric shape to avoid the external electromagnetic backgrounds. The $B-L$
gauge potential source
at a distance of $d$ is
made of Graphene. 
}
\label{fig:Design}
\end{figure}

{\bf Projected Constraint} --
For the $B-L$ gauge boson mass
$m_{A'} \lesssim 10^{-2}\,$eV,
the gauge coupling $g_{B-L}$ has already been strongly 
constrained from the fifth force detection experiments, 
like torsion balance \cite{Wagner:2012ui,Adelberger:2009zz}, and gravitational wave detection \cite{LIGOScientific:2021ffg,Miller:2023kkd,Frerick:2023xnf}. 
Meanwhile, the range $m_{A'} \gtrsim 1\,$eV is also constrained by stellar cooling effect \cite{Hardy:2016kme}.
Even in the mass range of $10^{-3}\,$eV$ \lesssim m_{A'} \lesssim 1\,$eV, the limit has reached as $g_{B-L} \lesssim 10^{-15}$. 
All the current constraints are shown 
as the color shaded region in \gfig{fig:limit}.
Thus, the strength of $B-L$ gauge potential is tiny.
Our experimental proposal compares the differential signals between two sides of JJ before and after the introduction of the plate, effectively canceling out the majority of environmental background.
Other backgrounds, including 
the Casimir effect, gravitational force, and 
the thermal noise, may still affect our experiment and we shall discuss them as follows.

\noindent \textbf{I. Background Electromagnetic Field:} 
The quantum 
fluctuation of field will induce some electromagnetic force between the 
surfaces of two 
macroscopic objects. This is the so-called Casimir force \footnote{Notice that the Van der Waals force can be neglected for macroscopic objects \cite{Klimchitskaya:2015vra}.}. 
The plate induces the Casimir force on the surfaces of the superconductors. 
However, this macro force does not act on the internal Cooper pairs and does not affect their phase \cite{Klimchitskaya:2015vra}.

In addition, the laboratory is also subject to a residual background electromagnetic field, such as the presence of the geomagnetic field.
Therefore, experiments should be conducted in environments with electromagnetic shielding. Current technology allows electromagnetic fields to be shielded to very low levels \cite{Westphal:2020okx}, and any remaining electromagnetic fields affecting both sides of the JJ equally will not impact the measurement of the phase difference.

\noindent \textbf{II. Gravitational Potential:} The plate made of Graphene not only 
produces the $B-L$ gauge potential, it is also a gravitational potential source. 
One carbon atom has a $B-L$ charge $Q_{C} = 6$ and a gravitational
charge $m_C \simeq 12\,$amu\,$\simeq 12\,$GeV. 
Interacting with an  electron, 
the gravitational force 
is the same order of $B-L$ gauge force, $G m_C m_e/ r^2 \simeq g_{B-L}^2 \times 6/r^2$ 
($G = 6.8 \times 10^{-39}\,$GeV$^{-2}$ is the Newton constant), 
when $g_{B-L} \simeq 10^{-20}$ for a simple estimation. 
Below that, the gravitational force dominates. For the same
settings, the quantum 
phase induced by gravitational potential
is, 
\begin{equation}
    \phi_g (d)
=
    \tau \times 
    \int_0^b
    \varoiint_{S}
    \frac{2 G m_e \rho_C}{\sqrt{(d+z)^2 + x^2 + y^2}} dS dz,
\end{equation}
where we take the surface area of the plate $S = 1\,$cm$^2$. After running for a minute, the phase difference induced by gravity is 
$\Delta \phi_g = \phi_g (d) - \phi_g (d + \delta) 
\simeq 10^{-11}$ for 
$d = 1\,\mu m$ and $b = 1\,$cm,
which is negligible
comparing with the \geqn{dphi2}. 
Furthermore,
the gravity force from the 
Earth can be neglected by 
tuning the height of two 
superconductors for equalizing 
the gravitational potentials in both sides.

\noindent \textbf{III. Thermal Noise:}
The thermal noise produces 
an inevitable background 
current, 
$I_T \approx e k T/ \hbar \approx 10^{-7} (T/1 K) A$ \cite{Perez:2023tld}. 
In the laboratory, researchers have achieved significantly low temperature 
$T < 0.1\,$mK 
in controlled environments \cite{Deppner:2021fks}. 
However, the main hurdle lies in replicating these conditions on a larger scale, both in terms of space and time. For a 
conservative estimation, 
we take the environment temperature $T = 1\,$mK
and the thermal current is $I_T = 10^{-10}\,$A.

\noindent \textbf{IV. Number-Phase Uncertainty:}
For a coherent state containing $N$ particles, its phase
has a quantum fluctuation due to the number-phase uncertainty relationship $\Delta N \cdot \Delta \phi \gtrsim \hbar$
\cite{Carruthers:1965zz}.
With the Cooper pair number density $n \simeq 1.2 \times 10^{22}$\,cm$^{-3}$ as mentioned above, the JJ of size $a^3$ has a total number of Cooper pairs $N \simeq 10^{7}$. 
The number fluctuation is then
$\Delta N \simeq \sqrt{N} \simeq 3 \times 10^{3}$.
As a result, the phase sensitivity can reach the level of $\Delta \phi \simeq 10^{-3}$.

In summary, the backgrounds from 
gravity and thermal noise are much 
smaller than the expected signal.
Therefore, we set the only condition for detecting the signal that the phase difference induced by the $B-L$ gauge force shall accumulate to surpass the typical threshold, $\Delta \phi_f > 10^{-3}$, as required by the number-phase uncertainty\cite{everitt2011gravity}.
Such a requirement gives a projected constraint for the $B-L$ gauge coupling
as the colorful lines in \gfig{fig:limit}.

Our proposal gives the strongest exclusion of the coupling $g_{B-L} \gtrsim 10^{-17}$ 
for F\'eeton mass $10^{-2}\,$eV\,$< m_{A'}<1\,$eV by assuming no signals 
detected. 
The limit becomes 
weaker when $m_{A'} \gtrsim 1/d$
since the Yukawa potential is
exponentially suppressed by distance. Thus, a smaller source
distance $d = 0.1\,\mu$m (red lines)
has a wider detectable range than 
the larger distance case $d = 1\,\mu$m
(orange lines). The distance of $\mathcal{O}(0.1)\,\mu$m is currently the smallest achievable gap between objects in experiments \cite{Decca:2011vua}.
On the other hand, in the long range force limit $1/b < m_{A'} < 1/d$,
the phase difference, 
or the constraint, is independent of $d$ and is inversely 
proportional to $m_{A'}$
as indicated by \geqn{phaseDiff}.
In addition, 
a longer running time naturally yields stronger experimental constraints.
If $\tau$
increase by a factor of $60$
from $1\,$minute (solid) to $1\,$hour (dashed), the limit stronger by a factor
of $\sqrt{60}$ correspondingly.
The time $\tau$ of pure quantum evolution without losses is taken as order of one minute for a conservative illustration \cite{JacksonKimball:2017elr,Wu:2019exd,Garcon:2019inh,Ummarino:2020loo,Perez:2023tld}.

{\bf Discussion and Conclusions} --
In this paper, we propose a novel 
experimental design
that utilizes a superconducting Josephson junction to detect the 
fifth force mediated by $B - L$ gauge boson. 
The Josephson junction consists of two superconductors at different distances from the $B-L$ gauge potential source. As a result, 
a potential difference arises between the two Cooper-pairs-coherent states in the superconductors, 
leading to the generation of a phase difference over time. Due to the Josephson effects, this phase difference can be converted into an electric current and detected. By extending the running time, the signal can be significantly amplified, enabling more precise observation of the fifth force.

After considering the background from number-phase uncertainty, 
the experiment has the strongest sensitivity to $g_{B-L} \subset (10^{-18},10^{-15})$, for the gauge 
boson mass $m_{A'} \subset (10^{-2}\,$eV$,10\,$eV)
To achieve this result, a reasonable running time of $\tau \sim 1\,$minute was considered.

It is the first time
that a theoretical proposal has been made for detecting a fifth force experimentally at such small scales
below millimeter.
This result is not limited to the 
$B-L$ extension and can be generalized to 
other fifth force searches. It also covers some of interesting parameter region of DM search.
We hope that the present experimental proposal for measuring fifth forces becomes a new avenue in searching for new physics.

\section*{Acknowledgements}
The authors thank Jiadu Lin, Zixuan Dai, Qingdong Jiang, and Shao-Feng Ge, Michimura Yuta  for useful discussions. 
And T.T.Y. thanks Kazuya Yonekura for a very helpful discussion on the gauge invariant description of the system.
This work is supported by 
the National Natural Science
Foundation of China (12175134, 12375101, 12090060, 12090064, and 12247141),
JSPS Grant-in-Aid for Scientific Research
Grants No.\,19H05810, 
the SJTU Double First Class start-up fund No.\,WF220442604,
and World Premier International Research Center
Initiative (WPI Initiative), MEXT, Japan.

\providecommand{\href}[2]{#2}\begingroup\raggedright\endgroup

\appendix
\section{Model for the $B-L$ extension}

In this paper, we consider a simple model that extends the SM with a $B-L$ gauge group. The related Lagrangian is, 
\begin{align}
\mathcal{L}
&=\frac{i}{2} \bar{N}_i \gamma^\mu \partial_\mu N_i
+
\left(\lambda_{i \alpha} \bar{N}_i L_\alpha H -\frac{1}{2} M_{R i} \bar{N}_i^c N_i+\text { h.c. }\right)
\nonumber \\
& - 
\frac{1}{2} g_{B-L} \bar{N}_i \gamma^\mu \gamma_5 N_i A_\mu^{\prime}+
g_{B-L} q_{B-L} \bar{f} \gamma^\mu f A_\mu^{\prime}.
\end{align}
Here, $L_\alpha$, $f$, $H$ and $N$
are the SM left-handed lepton
doublets, fermions, Higgs boson and RHNs. The index $i$ stands 
for the generation of RHN and 
$\alpha$ for all the species of leptons. 
A new Higgs field $\Phi$ is introduced to break the gauged $B-L$ symmetry spontaneously and 
the vacuum expectation value of $\Phi$ is the breaking scale of 
$B-L$ gauge symmetry $\langle \Phi \rangle = V_{B-L}$. Its couplings to RHNs $N_i (i=1-3)$ generate large Majorana masses for them through 
$\frac{1}{2} h_i \Phi N_i N_i$.
Thus, the RHN mass is $M_{Ri} = h_i V_{B-L}$.
The $B-L$ quantum numbers 
$q_{B-L}$ in the present model for all particles are shown in \gtab{tab:QBL}. 

\begin{table}[h]
\centering
\caption{$B-L$ charge for different species}
\label{tab:species}
\begin{tabular}{|c@{\hspace{8pt}}|c@{\hspace{8pt}}|c@{\hspace{8pt}}|c@{\hspace{8pt}}|c@{\hspace{8pt}}|c@{\hspace{8pt}}|c@{\hspace{8pt}}|c@{\hspace{8pt}}|c@{\hspace{8pt}}|}
\hline
{\rm Species} & \centering $q_\alpha$ & $u_R$ & $d_R$ & $L_\alpha$ & $e_R$ & $N_i$ & $\Phi$ & $H$ \\
\hline
$q_{B - L}$ & 1/3 & 1/3 & 1/3 & -1 & -1 & -1 & 2 & 0\\
\hline
\end{tabular}
\label{tab:QBL}
\end{table}
The $q_\alpha$, $u_R$, $d_R$ and $e_R$ are the left-handed quark doublets, right-handed up- and down-type quarks, and the right-handed charged leptons, respectively. 
Regarding composite states, a proton or neutron is made up of three quarks with $B-L$ charge $+1$. Furthermore, the $B-L$ charge of an atom equals to 
the number of neutrons, since
the charges of electrons always
cancel out that of the protons.

\vspace{15mm}
\end{document}